\documentclass[prb,twocolumn,groupedaddress]{revtex4}
\usepackage{graphicx}
\usepackage{dcolumn}
\usepackage{bm}

\begin{document}

\title{Mode Confinement in Photonic Quasi-Crystal\\
Point-Defect Cavities for Particle Accelerators}

\author{Emiliano Di Gennaro}
\email[]{emiliano@na.infn.it}
\affiliation{CNISM and Department of Physics, University of Naples ``Federico II'', I-80125 Naples, Italy}

\author{Salvatore Savo}
\affiliation{CNISM and Department of Physics, University of Naples ``Federico II'', I-80125 Naples, Italy}

\author{Antonello Andreone}
\affiliation{CNISM and Department of Physics, University of Naples ``Federico II'', I-80125 Naples, Italy}

\author{Vincenzo Galdi}
\affiliation{Waves Group, Department of Engineering, University of
Sannio, I-82100 Benevento, Italy}

\author{Giuseppe Castaldi}
\affiliation{Waves Group, Department of Engineering, University of
Sannio, I-82100 Benevento, Italy}

\author{Vincenzo Pierro}
\affiliation{Waves Group, Department of Engineering, University of
Sannio, I-82100 Benevento, Italy}

\author{Maria Rosaria Masullo}
\affiliation{I.N.F.N., Unit of Naples, I-80125 Naples, Italy}

\date{\today}

\begin{abstract}
In this Letter, we present a study of the confinement properties of
point-defect resonators in finite-size photonic-bandgap structures
composed of aperiodic arrangements of dielectric rods, with special
emphasis on their use for the design of cavities for particle
accelerators. Specifically, for representative geometries, we study
the properties of the fundamental mode (as a function of the filling
fraction, structure size, and losses) via 2-D and 3-D full-wave
numerical simulations, as well as microwave measurements at room
temperature. Results indicate that, for reduced-size structures,
aperiodic geometries exhibit superior confinement properties by
comparison with periodic ones.
\end{abstract}
\maketitle

A promising approach to the development of new types of compact,
efficient microwave particle accelerators is to realize structures
having higher frequencies of operation, since this can increase the
accelerating field gradient and reduce the power consumption
\cite{Whittum}. Unfortunately, in a particle accelerator, the higher
the frequency, the stronger the excitation (by the wakefield effect)
of higher order modes (HOMs), with a significant reduction of the
beam stability. Nowadays accelerators operate at frequencies
$\lesssim 1$ GHz, and they usually rely on waveguide-based HOM
dampers in order to efficiently suppress the wakefields. However, at
higher ($\gtrsim$ 10 GHz) working frequencies, the standard
configurations used for HOM damping become rather cumbersome or even
technically unfeasible.

In the past, open metallic photonic-crystal (PC) cavities, based on
periodic arrangements of inclusions, have been proposed as
candidates for a new generation of accelerating cells since they
exhibit electromagnetic (EM) responses that can be highly selective
in frequency. This property (and others related) arises from the
formation of photonic bandgaps (PBGs), whereby multiple scattering
of waves by periodic lattices of inclusions acts to prevent the
propagation of EM waves within certain frequency ranges. In these
structures, an open resonator can be easily created by introducing a
lattice {\em defect}, e.g., by removing one inclusion. Unlike
conventional closed cavities, such a {\em point-defect} PBG cavity
can be designed to support only one bound mode (strongly {\em
localized} within the defect region), and  {\em mostly extended}
HOMs. This suggests the possibility of using PBG-based cavities for
effective HOM wakefield suppression, without the need for mode
couplers or detuning. Indeed, a prototype of a large-gradient
accelerator that relies on a metallic PBG structure has been
successfully fabricated and experimentally characterized
\cite{Smirnova1}. Superconducting prototypes have been also tested
at 4 K, showing quality factors up to $\sim 10^5$, limited by
radiation losses only \cite{Andreone}. {\em All-dielectric} or {\em
hybrid-dielectric} PBG structures could also be used, thereby
eliminating or reducing the characteristic metallic losses at the
frequency of operation, even if at the expense of an increased
radiative contribution \cite{Masullo}.

Moreover, it should be observed that spatial periodicity is not
an essential ingredient for obtaining PBG and related confinement
effects. During the past few years, following up on the discovery of
``quasicrystals'' in solid-state physics \cite{Senechal},
similar effects have been found in {\em aperiodically-ordered}
structures too (see, e.g., Ref. \onlinecite{Steurer} for  a recent
review). Such structures, commonly referred to as ``photonic
quasicrystals'' (PQCs), are typically based on the so-called
``aperiodic-tiling'' geometries \cite{Senechal}, characterized by
weak (local or statistical) rotational symmetries of
``noncrystallographic'' type (e.g., of order 5, 8, 12).  In PQCs,
the EM response can be strongly dependent on the lattice short-range
configuration and on multiple interactions \cite{DellaVilla}, and
yet almost independent of the incidence angle, and can be
engineered/optimized via judicious exploitation of the additional
degrees of freedom endowed by aperiodicity (see, e.g., Ref.
\onlinecite{Wang1}). Moreover, in view of their many non-equivalent
sites, PQCs exhibit a number of possible different defects useful
for field confinement\cite{Chan}, as well as intrinsic localized
modes \cite{Wang,DellaVilla2}. In this connection, PQC-based optical
microcavities have already been successfully fabricated, with the
ability of strongly confining the light with both high quality
factors and small modal volumes \cite{Nozaki,Lee}.

In this Letter, we present a comparative study of the confinement
properties of different point-defected PBG cavities made of
cylindrical dielectric rods arranged according to representative
aperiodic PQC geometries. The study is aimed at highlighting the
potential advantages offered by PQC dielectric structures in the
design of PBG-based accelerating resonators, with respect to their periodic PC
counterparts. In this framework, we present a body of
results from numerical full-wave simulations to compare the
confinement properties of the various configurations (including a reference
periodic one) as a function of the structure size,
filling fraction, and losses. In addition, for experimental
verification, we present the results of measurements at room
temperature on cavity prototypes operating in the microwave region
($\sim 16.5$ GHz).

\begin{figure}[hbtp!]
\centering
\includegraphics[width=0.495\textwidth]{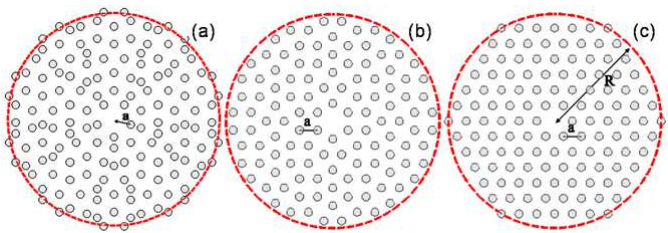}
\caption{(color online) Point-defect PBG cavity geometries for Penrose (a),
dodecagonal (b), periodic (triangular) (c) lattices, with structure
radius $R=7a$ and filling fraction $r/a=0.2$.}\label{hex_dod}
\end{figure}

The structures of interest are composed of sapphire (relative
permittivity $\varepsilon_r$ = 9.2, and typical loss-tangent of
$10^{-6}$) circular rods of radius $r=0.15$ cm and height $h=0.6$
cm, placed at the vertices of a periodic or aperiodic tiling with
lattice constant $a=0.75$ cm. The structure is sandwiched between
two metallic plates, and presents a bore radius of $0.2$ cm for the
particle beam transit. Specifically, as shown in Fig. \ref{hex_dod},
we consider two representative (aperiodic) PQC geometries based on
the Penrose (5fold-symmetric, Fig. \ref{hex_dod}(a)) and on the
dodecagonal (12fold-symmetric, Fig. \ref{hex_dod}(b)) tilings (see,
e.g., Refs. \onlinecite{Senechal,Oxborrow} for details about their
generation), and a reference periodic (triangular) PC geometry (Fig.
\ref{hex_dod}(c)).

The structures are obtained by cutting, from suitably large lattices
with same rod properties and lattice constant, a circular area of
radius $R$, and finally removing the central rod, as shown in Fig.
\ref{hex_dod}, in coincidence with the beam transit aperture. We
assume time-harmonic ($\exp(-i2\pi f t)$) excitation and transverse
magnetic (TM) polarization (electric field parallel to the rods
axis).

Our full-wave numerical study of the EM response of the structures
is based on the combined use of a commercial 3-D simulator (CST Microwave Studio) based on
finite integration in the time domain and an in-house 2-D
code based on a Bessel-Fourier multipole expansion \cite{Tayeb}.
The two main loss mechanisms are related to radiation (in view of
the open character of the structure) and Ohmic dissipation in the
metallic plates, the sapphire dielectric losses being actually
negligible. Assuming the metallic plate separation smaller than half
a wavelength, so that the field distribution is rather uniform in
the direction along the rods axis, the radiation losses can be
efficiently modeled using the 2-D code (which assumes the dielectric
rods infinitely long). This renders a parametric study
computationally affordable.

\begin{figure}[hbtp!] \centering
\includegraphics[width=0.43\textwidth]{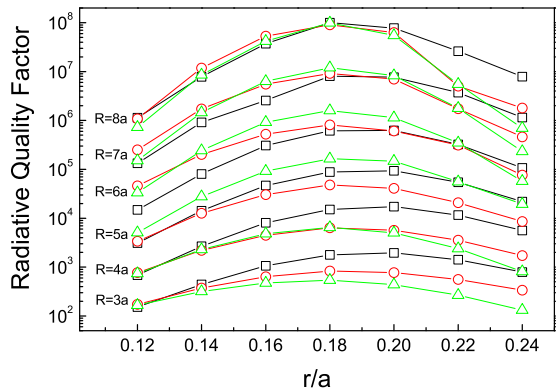}
\caption{(color online) Radiative quality factor of the fundamental TM mode, for
Penrose (triangles), dodecagonal (squares), and periodic (circles)
geometries, as a function of the filling fraction $r/a$, and for
different values of the structure radius $R$.}\label{QvsR}
\end{figure}

\begin{figure}[hbtp!]
\includegraphics[width=0.495\textwidth]{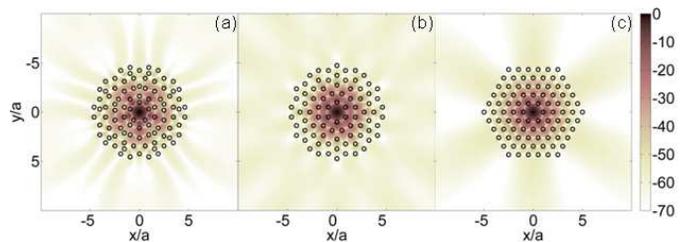}
\caption{(color online) Normalized electric field intensity maps (in dB) of the
fundamental resonant TM mode for Penrose (a), dodecagonal (b), and
periodic (c) configurations, with $R=5a$ and $r/a=0.2$.}\label{att}
\end{figure}

Figure \ref{QvsR} shows, for the three geometries of interest, the
radiative quality factor ($Q_{r}=f_c^{(r)}/2f_c^{(i)}$, with
$f_c^{(r)}+if_c^{(i)}$ denoting the complex eigenfrequency obtained
by solving a source-free problem \cite{Tayeb}) dependence on the
filling fraction $r/a$ and the structure size $R$. We observe that,
for all geometries, a maximum $Q_{r}$ is achieved for $r/a$ ranging
between 0.18 and 0.20. Differences among the geometries emerge when
considering the dependence on the structure size $R$. Specifically,
dodecagonal PQC cavities outperform the periodic ones for small to
moderate structure sizes, and become eventually comparable for
$R\gtrsim8a$. Particularly intriguing is the behavior of the Penrose
PQC cavities, which exhibit the lowest $Q_{r}$ for smaller sizes
($R\lesssim 4a$), and the highest for moderate cavity sizes. Similar
indications emerged from the study of the electric field spatial
distribution of the resonant modes, shown in Fig. \ref{att} for a
moderate structure size $R=5a$, so as to better visualize the power
radiated outside the cavity.

As compared with the periodic case, radiation in PQC structures
appears to be much better confined. In the case of the dodecagonal
structure, the field configuration is also more isotropic than in
the periodic case, due to its higher rotational symmetry. Results
tend to become much more similar for larger structures.

We also studied other tiling geometries (e.g., octagonal
\cite{Senechal, Chan}), whose responses, however, turned out not to
be particularly interesting (being always intermediate between the
dodecagonal and periodic ones), and are accordingly not shown.

For modeling the conducting losses of the metal plates, which become
dominant for larger-size structures, we use the
commercial (CST Microwave Studio) 3-D simulator , whose eigenmode solver option
also allows the computation of all the main parameters (shunt
impedance, conducting and dielectric loss contributions) of interest
for accelerating cavities, with the exception of the radiation
losses.

\begin{table}[!htpb]
\caption{Simulated characteristic parameters for the fundamental TM
mode of point-defect mid-sized cavities ($R=4a$,$5a$) having Penrose
(PEN), dodecagonal (DOD), and periodic triangular (PER) geometries,
and filling fraction $r/a=0.2$}\label{tabQ}

\begin{tabular}{ | c |c c c | c c c|}\hline
&&$R=4a$&&&$R=5a$&\\\cline{2-7}
 &PEN&DOD&PER&PEN&DOD&PER \\\hline\hline
$f_c [GHz]$&16.788&16.424&16.556 &16.770&16.429&16.561\\
$Q_c$&11400&11800&11900 &12200&11900&11900 \\
$Q_{r}$&5000&17200&5700 &147000&93000&40800\\
$r_{s}[M\Omega/m]$ &60&180&100 &280&280&240  \\
\hline
\end{tabular}

\end{table}

Table \ref{tabQ} summarizes, for the cases under test and for the
fundamental TM mode, the resonance frequency, the Ohmic and
radiative quality factors $Q_c$ and $Q_r$ respectively, and the
shunt impedance per unit length $r_{s}$, defined as $V_o^2/(2Ph)$,
where $V_o$ is the voltage ``seen'' by the incoming particle and $P$
the EM power lost in the PBG cavity. In the simulation analysis,
conducting losses are computed at $300K$ assuming the metal plates
as made of high-purity copper (electrical conductivity $\sigma$ =
$5.8\cdot 10^7$ $S/m$).

The $r_{s}$ values well compare with data previously reported on PBG
metallic cavities \cite{Smirnova1}, and suggest that hybrid
dielectric structures can be successfully exploited for the design
of high-gradient accelerators.

\begin{figure}[!htpb]
\centering
\includegraphics[width=0.45\textwidth]{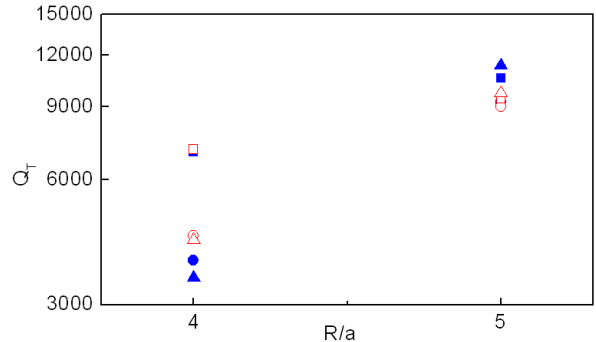}
\caption{(color online) Experimental (open symbols) and simulated (full symbols)
unloaded quality factor $Q_T$ values for Penrose (triangles),
dodecagonal (squares), and periodic (circles) lattices, having
filling fraction $r/a=0.2$, and structure radius $R=4a$ and
$R=5a$.}\label{Qt}
\end{figure}

For experimental verification, we fabricated prototypes of the
simulated structures, by suitably placing single crystal sapphire
rods between two Oxygen-Free High-Conductivity (OFHC) copper plates.
In the experimental setup, input and output tiny dipole antennas are
placed on the top plate; the input antenna feeds the cavity near
(but non in correspondence of) the defect region, so as to avoid
extreme reflection from that port, whereas the output antenna is
sufficiently far from the dielectric rods. Both input and output
antennas are then connected to a HP8720C Vector Network Analyzer.
Careful attention is paid to ensure that measurements are carried
out in a weak coupling configuration (insertion losses at the
resonance $\sim$ - 40dB). From the frequency response of the
transmission scattering parameters, we evaluate the resonance
frequency $f_c$ and the unloaded overall quality factor $Q_T$
for mid-size structures ($R/a$ = 4, 5) featuring each of the
geometries under study. Concerning the spectral response, results
confirm the intrinsically monomodal behavior of PQC cavities, with
the defect mode being the only one transmitted well within the
corresponding bandgap. The measured resonance frequencies agree
pretty well (within $2\%$) with the numerical predictions in Table
\ref{tabQ}. There is also a good agreement between the
experimentally measured $Q_T$ and the values extracted from
simulations ($Q_T=(Q_r^{-1}+Q_c^{-1})^{-1}$), as shown in Fig.
\ref{Qt}.

For $R=4a$, the performance is limited by radiation losses only, and
therefore the $Q_T$ factor is strongly dependent on the geometry, as
expected from the results shown in Fig. \ref{QvsR}. Surprisingly,
the measured quality factors are always higher than those predicted
by the simulations. Conversely, for $R=5a$, $Q_T$ is almost
independent of the dielectric rods arrangement, since it is strongly
dominated by the conductive losses. The slight discrepancy between
measurements and simulations in this latter case should be ascribed
to the inherent uncertainty in the copper conductivity values
considered in the modeling (see, e.g., Ref.
\onlinecite{Inagaki}).

It is worth mentioning that the quality factors of our
hybrid-dielectric aperiodic structures are higher than those found
for metallic periodic cavities ($\sim 4\cdot 10^3$) of similar size
and resonance frequency ($\sim 17.1$ GHz)\cite{Smirnova1}. This is
due to the combined reduction of both Ohmic and radiative losses, by
using dielectric inclusions and exploiting aperiodic geometries.
Additional advantages of hybrid-dielectric resonators include a
reduced complexity in the fabrication, and the possibility of
operating at higher frequencies using HOMs without mode competition
\cite{Masullo}. Moreover, it is worth noting that, in principle,
breakdown and charging can be strongly reduced or avoided in a
dielectric accelerating structure \cite{Gai, Hill}.

To sum up, our comparative study indicates that, although in the
limit of very large structure size the confinement properties of
periodic and aperiodic geometries tend to become comparable, for
small to moderately-sized structures specific aperiodic
configurations (in our case, Penrose and dodecagonal) turn out to
outperform the reference periodic one. In particular, for the
structure size $R=5a$, we found the Penrose-type geometry to provide
a considerable improvement (of nearly a factor 4) in the radiative
quality factor, as compared with the periodic (triangular)
counterpart (see Table \ref{tabQ}). The physical interpretation of
such observation can be addressed recalling the mechanisms
underlying the bandgap formation in PQCs, which, at variance with
the periodic PC case, are not necessarily related to long-range
interactions. With specific reference to Penrose-type PQCs, it was
shown in Ref. \onlinecite{DellaVilla} that the bandgap pertaining to
the point-defect mode of interest here is actually attributable to
{\em short-range} interactions. Accordingly, for reduced-size
structures, we intuitively expect such a geometry to provide a more
effective field confinement than a periodic counterpart.

An improvement in the quality factor, achievable via a simple
spatial rearrangement of the inclusions, constitutes indeed an
attractive perspective toward the design of cavities for compact,
efficient microwave particle accelerators. However, as shown by our
numerical and experimental results, such improvement can actually be
washed out by conducting losses in the metallic plates. In this
framework, current and future studies are aimed at the fabrication
and testing of cryogenic hybrid-dielectric prototypes. Preliminary
results show an improvement between 2 and 3 in the overall quality
factor, with decreasing the operating temperature to $\sim 100$ K,
mainly due to the reduction of the conducting losses. The use of
{\em superconducting} plates could be also foreseen, along the lines
of the results presented in Ref. \onlinecite{Andreone}. In this last
case, however, a deeper study should be undertaken in order to
understand to what extent the improvement in the cavity performance
would justify the higher complexity introduced by the necessity of
operating at 4 K (or even less).

The authors wish to thank Mr. C. Zannini for carrying out the CST simulations.
This work was supported in part by the Italian Ministry of Education and
Scientific Research (MIUR) under a PRIN-2006 grant, and in part by the
Campania Regional Government under a 2006 grant L.R. N. 5 -- 28.03.2002.

\end{document}